\documentclass[sigconf]{acmart}
\AtBeginDocument{%
  \providecommand\BibTeX{{%
    \normalfont B\kern-0.5em{\scshape i\kern-0.25em b}\kern-0.8em\TeX}}}

\copyrightyear{2024}
\acmYear{2024}
\setcopyright{acmlicensed}\acmConference[ICSE '24]{2024 IEEE/ACM 46th International Conference on Software Engineering}{April 14--20, 2024}{Lisbon, Portugal}
\acmBooktitle{2024 IEEE/ACM 46th International Conference on Software Engineering (ICSE '24), April 14--20, 2024, Lisbon, Portugal}
\acmPrice{15.00}
\acmDOI{10.1145/3597503.3608136}
\acmISBN{979-8-4007-0217-4/24/04}

\acmConference[ICSE 2024]{46th International Conference on Software Engineering}{April 2024}{Lisbon, Portugal}
\usepackage{subfig}
\usepackage{multirow}
\usepackage{tcolorbox}
\usepackage{balance}
\DeclareMathOperator*{\argmax}{arg\,max}

\begin{document}

%\title{FAIR: Flow Type-Aware Code Representation Pre-Training via Compiler Intermediate Representation}
\title{FAIR: Flow Type-Aware Pre-Training\\ of Compiler Intermediate Representations}
\renewcommand{\shorttitle}{FAIR: Flow Type-Aware Pre-Training of Compiler Intermediate Representations}

% \affiliation{%

% \affiliation{%

\author{Changan Niu}
\affiliation{
  \institution{State Key Laboratory for Novel Software Technology\\Nanjing University}
  % \streetaddress{22 Hankou Road}
  \city{Nanjing}
  \country{China}}
\email{niu.ca@outlook.com}

\author{Chuanyi Li}
\affiliation{
  \institution{State Key Laboratory for Novel Software Technology\\Nanjing University}
  % \streetaddress{22 Hankou Road}
  \city{Nanjing}
  \country{China}}
\email{lcy@nju.edu.cn}

\author{Vincent Ng}
\affiliation{
  \institution{Human Language Technology Research Institute\\University of Texas at Dallas}
  % \streetaddress{Richardson, TX 75083-0688}
  \city{Richardson}
  \state{Texas}
  \country{USA}}
\email{vince@hlt.utdallas.edu}

\author{David Lo}
\affiliation{
  \institution{School of Computing and Information Systems\\Singapore Management University}
  \country{Singapore}}
\email{davidlo@smu.edu.sg}

\author{Bin Luo}
\affiliation{
  \institution{State Key Laboratory for Novel Software Technology\\Nanjing University}
  % \streetaddress{22 Hankou Road}
  \city{Nanjing}
  \country{China}}
\email{luobin@nju.edu.cn}


\begin{abstract}
While the majority of existing pre-trained models from code learn source code features such as code tokens and abstract syntax trees, there are some other works that focus on learning from compiler intermediate representations (IRs). Existing IR-based models typically utilize IR features such as instructions, control and data flow graphs (CDFGs), call graphs, etc. However, these methods confuse variable nodes and instruction nodes in a CDFG and fail to distinguish different types of flows, and the neural networks they use fail to capture long-distance dependencies and have over-smoothing and over-squashing problems. To address these weaknesses, we propose FAIR, a \textbf{F}low type-\textbf{A}ware pre-trained model for \textbf{IR} that involves employing (1) a novel input representation of IR programs; (2) Graph Transformer to address over-smoothing, over-squashing and long-dependencies problems; and (3) five pre-training tasks that we specifically propose to enable FAIR to learn the semantics of IR tokens, flow type information, and the overall representation of IR. Experimental results show that FAIR can achieve state-of-the-art results on four code-related downstream tasks.
\end{abstract}

\maketitle

\section{Introduction}
\label{section:introduction}

Recent years have seen the dramatic development and tremendous success of pre-trained models of code, such as CodeBERT~\cite{feng2020codebert}, GraphCodeBERT~\cite{guo2021graphcodebert}, PLBART~\cite{ahmad2021plbart}, CodeT5~\cite{wang2021codet5} and UniXcoder~\cite{guo2022unixcoder}. These pre-trained models employ code features such as code token sequences, abstract syntax trees (ASTs), and data flow graphs, and have achieved remarkable results on a variety of software engineering (SE) downstream tasks such as code summarization~\cite{feng2020codebert,ahmad2021plbart,wang2021codet5,guo2022unixcoder}, code search~\cite{feng2020codebert,guo2021graphcodebert,guo2022unixcoder}, code-to-code retrieval~\cite{feng2020codebert,guo2022unixcoder}, and defect detection~\cite{wang2021codet5}. While the overwhelming majority of SE researchers are working on the source code of high-level programming languages to explore better representation methods, others set their sights on compiler intermediate representations (IRs).

IR is a low-level representation of code used by the compiler infrastructure. Though described as ``low-level'', IR retains rich semantic information and can express high-level ideas. First, compared to other low-level languages such as the assembly language, IR is much easier to understand due to its higher-level abstraction, consistent syntax, platform independence, and register-agnostic. Moreover, IR is programming language-independent, which could provide a more concise, uniform, and efficient representation of code than a high-level programming language. However, the IR-based model requires a high demand on the dataset and needs the code to be compilable. Most of the current code datasets are collected with the compilation-related information removed, which implies that the code in them cannot be compiled. Consequently, compared with source code-based models, IR-based models are much less studied. Nevertheless, due to the unique advantages that IR has, IR-based research continues to flourish.

In existing IR representation learning research, IR features such as token sequences~\cite{peng2021oscar}, control flow graphs (CFGs)~\cite{venkatakeerthy2020ir2vec,yu2020codecmr}, control and data flow graphs (CDFGs)~\cite{ben2018ncc,brauckmann2020gnnast,cummins2021programl} are commonly used. As for the model architecture, existing work mostly chooses message-passing paradigm-based graph neural networks (GNNs) to encode graphical features ~\cite{brauckmann2020gnnast,cummins2021programl,yu2020codecmr}. Existing approaches also use other methods to learn the representation of an IR program. For example, in order to obtain the embedding vector of an IR instruction, inst2vec~\cite{ben2018ncc}, a skip-gram model, and seed embeddings~\cite{venkatakeerthy2020ir2vec}, a TransE~\cite{bordes2013transe} model, are trained on CDFs and CFG priors, but the key drawback of these approaches is that the resulting pre-trained embeddings are not task-agnostic and therefore cannot embed the contextual information of a target downstream task.

Nevertheless, there are several weaknesses in existing work (except for those based on GNNs) on IR-based models w.r.t.\ the IR features used by these models. Recall that in a CDFG, there are two types of nodes, one for variables/values (operands) and one for instructions, Existing approaches fail to distinguish between these two types of nodes by embedding in the same representation space using the same embedding method, while other approaches simply eliminate one type of nodes, which might greatly reduce performance~\cite{ben2018ncc,brauckmann2020gnnast}. In addition, existing work treats all flows as equivalent~\cite{yu2020codecmr} or does not completely distinguish between all flow types~\cite{brauckmann2020gnnast,ben2018ncc,cummins2021programl}, However, the flows in a CFG and a DFG should not be treated as identical. For example, in a CFG, a node may have multiple jump relationships controlled by conditions such as a Boolean expression, while in a DFG, the dependencies between data can be additive, divisive, etc. In fact, the flow-type information does exist in the original CDFG. For example, the flow type of a CFG can be retrieved from the last instruction of the basic block node, and the flow type of a DFG is in the opcode of the instruction node. Existing approaches embed the nodes first and then learn the flow information. As a result, the flow-type information stored in the nodes will be diluted by other texts in the nodes when performing node embedding, and more importantly, this flow-type information cannot be correctly associated with the corresponding flows in a CDFG.

Another weakness associated with existing work on IR-based models lies in the model architecture. Specifically, while existing work typically chooses message-passing-based GNNs to encode graphical features such as CDFGs, a CDFG is usually very large, often with more than a thousand nodes and thousands of flows. Such a large and densely connected graph would cause long-range dependencies~\cite{liu2021eignn,rampavsek2022recipe} problems for GNNs. Besides, the training process of GNNs naturally has over-smoothing and over-squashing problems, where the former refers to a situation where the representations of nodes become too similar to each other as a result of repeated graph convolutions~\cite{Oono2020Graph,chen2020measuring,kreuzer2021rethinking}, and the latter refers to a situation where the activation function used in the GNN model compresses the node representations too much, causing the model to lose important information~\cite{alon2021on,topping2022understanding}.

All things considered, there is no existing work that seeks to address the size and heterogeneity (i.e., different node/flow types) problem of CDFGs, as well as the problems caused by GNNs. In light of these observations, we propose FAIR, a \textbf{F}low type-\textbf{A}ware code pre-trained model based on \textbf{IR}. FAIR distinguishes itself from existing IR-based pre-trained models in its \emph{input representation}, \emph{model architecture}, and \emph{pre-training} tasks, as described below:

\emph{Input Representation.} FAIR (1) decomposes a CDFG into a CFG and a DFG in order to reduce graph size; (2) assigns an explicit Flow Type to each flow in both the CFG and the DFG to distinguish different flow types; (3) adds the flows according to the call graph in order to connect multiple CFGs or DFGs of one single IR program; and (4) adds flows to link the nodes from the CFG and those from the DFG that have reference relationships. This process yields a novel graph-based input representation of an IR program.

\emph{Model Architecture.}
FAIR (1) uses a Transformer Encoder~\cite{vaswani2017transformer} and a normal word embedding layer to embed the nodes of CFG and DFG, respectively; (2) employs Graph Transformer~\cite{dwivedi2020generalization} to learn the representation of the entire IR program by taking the nodes' embedding as input and injecting graph priors into the attention computation via graph bias terms; and (3) associates each flow type with a unique bias term in order to learn from flow types.

\emph{Pre-Training.} FAIR employs 
five pre-training tasks: (1) Masked Language Modeling (MLM)~\cite{devlin2019bert}, which enables the model to predict the original nodes in the CFG and  the DFG that are masked in the input; (2) 
CFG Flow Type Prediction (CFT), (3) DFG Flow Type Prediction (DFT), and (4) BB-Var Flow Prediction (BVP), all of which randomly mask some flows in the graph and then let the model predict whether these flows exist, and/or the flow type; and (5) a pre-training task based on contrastive learning, where we design four novel strategies to construct positive examples.

We compare FAIR with strong baselines based on both IR and source code on four downstream tasks, namely code-to-code retrieval, algorithm classification, heterogeneous device mapping, and optimal thread coarsening factor. Empirical results show that FAIR achieves state-of-the-art performance on all tasks
and generalizes very well to unseen programming languages.\footnote{Artifacts are available at \url{https://github.com/NougatCA/FAIR}.}

Overall, we make the following contributions. First, we propose FAIR, a flow type-aware pre-trained model of IR, which is programming language- and platform-independent. FAIR is novel in its design of an \emph{input representation} of IR programs as well as \emph{pre-training tasks} that aim to predict concrete types of flows and novel strategies to generate more positive examples for contrastive learning. Second, when pre-training FAIR on several large open-source repositories, we achieve state-of-the-art performance on four downstream tasks.

\section{Related Work}
\label{section:related}

\subsection{Source Code-based Pre-Trained Models}
\label{section:related_pre_train}

Inspired by the successes of pre-trained models in natural language processing (NLP), e.g., BERT~\cite{devlin2019bert}, RoBERTa~\cite{liu2019roberta}, BART~\cite{lewis2020bart} and T5~\cite{raffel2020t5}, a number of pre-trained models of source code have been proposed~\cite{feng2020codebert,guo2021graphcodebert,ahmad2021plbart,wang2021codet5,mastropaolo2021t5-learning,niu2022sptcode,guo2022unixcoder}. While some of the pre-training tasks used in these models are directly copied from NLP such as MLM and replaced token detection~\cite{liu2019roberta}, other pre-training tasks are designed to encode code content. In particular, code token-aware and natural language-aware pre-training tasks are widely adopted. For instance, identifier MLM only masks identifiers in the code tokens and trains the model to predict them~\cite{liu2020cuglm,roziere2021dobf,wang2021codet5}, and cross-modal generation~\cite{wang2021codet5,guo2022unixcoder} aims to generate natural language/code given code/natural language. Structure-aware pre-training tasks have also been proposed to enable a model to learn the structural information in, for instance, ASTs and DFGs. Examples include edge prediction and node alignment tasks, which help a model learn features within a DFG and between a DFG and code~\cite{guo2021graphcodebert}.

Contrastive learning is frequently used to improve the overall representation capability of a model. Existing contrastive learning strategies differ primarily in the methods used to generate positive examples. These methods include swapping the order of input parts, inputting different modalities of the same example separately~\cite{wang2021syncobert}, and using different dropout masks~\cite{guo2022unixcoder}.

Despite the successes of source code-based pre-trained models, we believe it is important to investigate IR-based models for at least two reasons. First, IR is programming language-independent, so IR-based models only need to capture the unique features of the IR language, such as grammar, vocabulary, and syntax. Second, IR-based models can be trained more efficiently since they do not require processing and aligning data from multiple languages.

\subsection{IR-based Models}
\label{section:related_ir}

Recent work on IR-based pre-trained models can be broadly divided into three categories:

\textbf{Using existing pre-trained models for node embedding.} Ncc~\cite{ben2018ncc} combine a Control Flow Graph (CFG) and a Data Flow Graph (DFG) in order to build a Contextual Flow Graph (XFG). With an XFG, they train inst2vec, a skip-gram-based pre-trained embedding lookup table for each IR instruction, by defining the context of size $N$ as nodes within distance $N$ in the XFG. Then, they use LSTM to verify the performance of the trained inst2vec on downstream tasks. IR2VEC~\cite{venkatakeerthy2020ir2vec} uses a trained embedding lookup table of seed embeddings. To obtain the lookup table, the authors (1) extract opcode, data type, and arguments from each instruction, (2) use the extracted information to convert an instruction into several triples, (3) apply the TransE learning model~\cite{bordes2013transe} to the resulting triples to learn the seed embeddings of each instruction. Based on seed embeddings, they add the information of a CFG to obtain the representation vectors of an IR program. Rather than utilizing a lookup table, Gui et al.~\cite{gui2022xlir} use a BERT model pre-trained on IR data to embed a given IR program.

\textbf{Using GNNs to encode graph features.} CodeCMR~\cite{yu2020codecmr} feeds the source code of a high-level language and the CFG of a low-level language into the DPCNN~\cite{johnson2017dpcnn} and the GNN, respectively. GNN-CDFG~\cite{brauckmann2020gnnast} (1) adds call graph and store-load dependencies into the CDFG of IR, (2) simplifies the nodes in the CDFG by eliminating the variable/value nodes and replaces each instruction node with its opcode, and (3) encodes the resulting graph using a message-passing paradigm-based GNN~\cite{li2016gated}. GNN-CEFG outperforms state-of-the-art approaches that use sequential models based on token sequences. ProGraML~\cite{cummins2021programl} (1) adds call graph to a CDFG and utilizes Message Passing Neural Network (MPNN) framework~\cite{gilmer2017mpnn} to encode the whole graph, and (2) uses opcode and data type to represent an instruction. Both of these work tries to address the heterogeneity of CDFG by discarding some critical information, such as operands and return values. However, the heterogeneous nature of CDFG is not considered~\cite{yu2020codecmr} or well handled~\cite{brauckmann2020gnnast,cummins2021programl}. Different from them, in FAIR, we decompose CDFG into CFG and DFG, and in addition to adopting a call graph, we define explicit types for flows, as well as simplify DFG and connect CFG and DFG with a novel type of flow.

\textbf{Developing pre-trained models of IR.} With the emergence of pre-training, some recent approaches utilize pre-training. OSCAR~\cite{peng2021oscar}, a pre-trained model of IR, leverages abstract environment information (AEI) along with the IR token sequence as model input. In contrast, IRGen jointly learns source code and the corresponding IR code generated using different compilation optimization options in order to better represent programs~\cite{li2022irgen}. As pre-training tasks, MLM is used by OSCAR, whereas contrastive learning is used by both OSCAR and IRGen, even though the way contrastive learning is being used is different in the two models. Specifically, to construct more positive examples for contrastive learning, OSCAR generates correct IRs for each source code with different compilation optimization options, whereas IRGen uses contrastive learning by extending CodeCMR with a new objective based on triplet loss that increases the similarity between a source code and its corresponding IR and at the same time reduces the similarity between the source code and the irrelevant IRs. While we also employ contrastive learning in the design of FAIR, we (1) propose four novel strategies to construct positive examples by mutating the input of the given IRs, and (2) design the other two novel pre-training tasks that had never been used by existing pre-training models of IR, i.e., predicting the flow type of CFG and DFG.

\begin{figure*}
    \centering
    \begin{minipage}[b]{.29\textwidth}
        \subfloat[An example of IR program.]
        {\includegraphics[width=\linewidth]{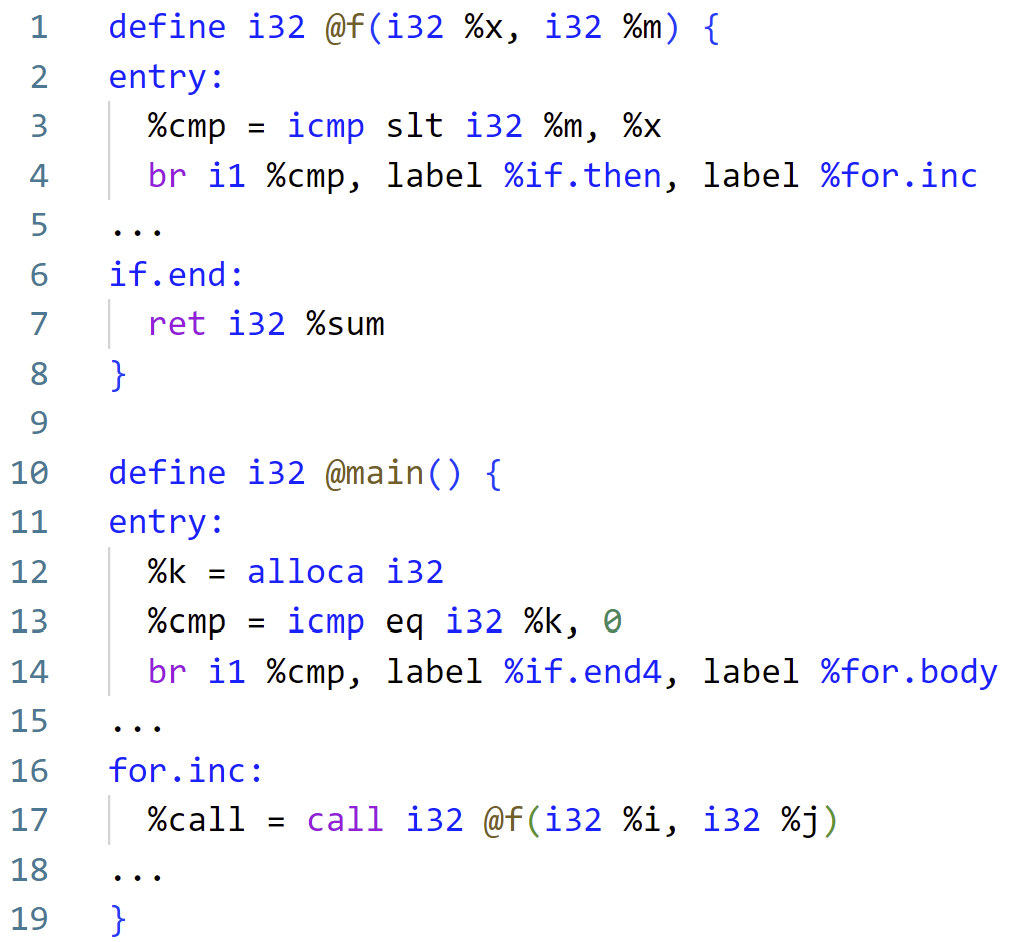}
        \label{figure:ir_example}}
    \end{minipage}\quad
  \begin{minipage}[b]{.32\textwidth}
  \subfloat[Adding flow type and call graph to CFG.]
    {\includegraphics[width=\linewidth]{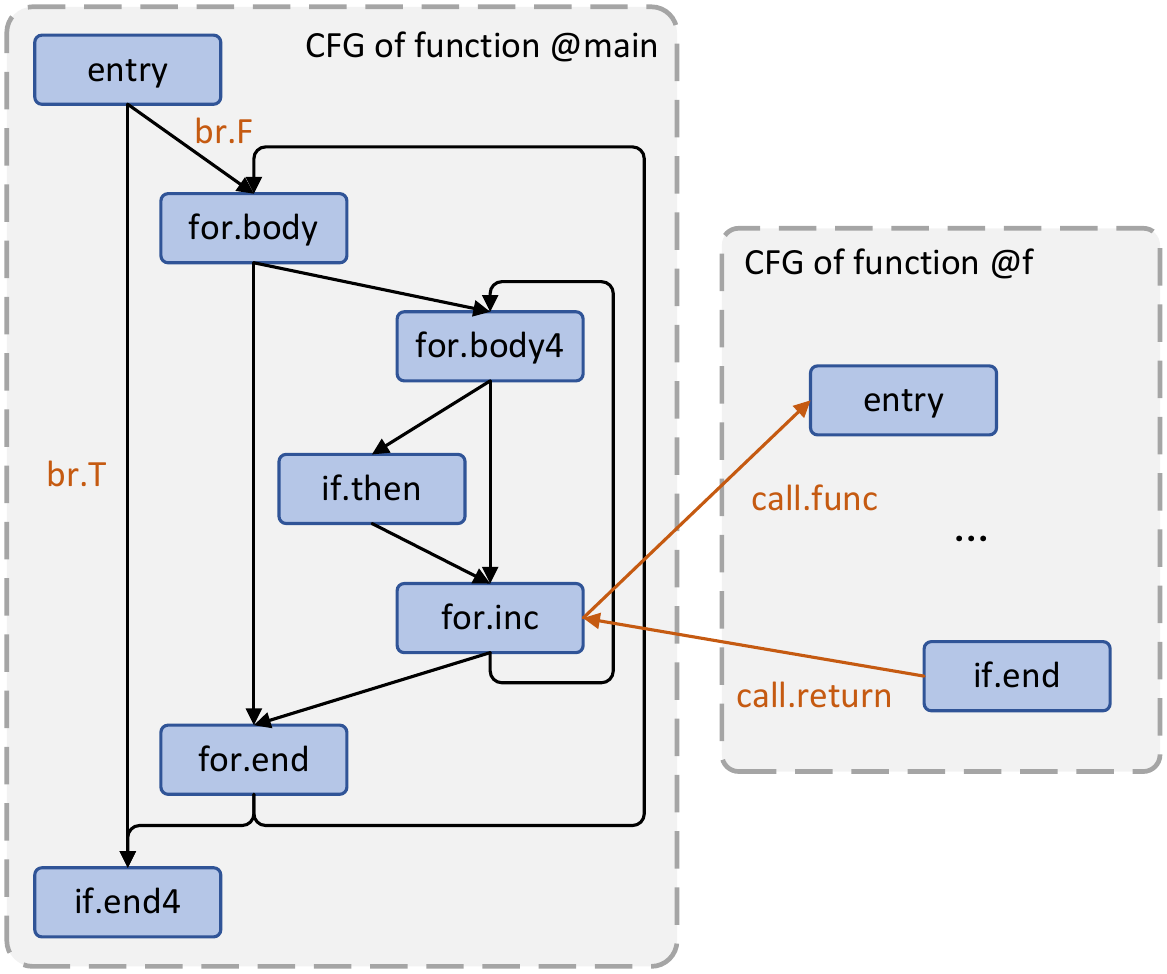}
    \label{figure:cfg}}
    \end{minipage}\quad
\begin{minipage}[b]{.29\textwidth}
\centering
  \subfloat[Simplifying DFG and adding flow type.]
    {\includegraphics[width=\linewidth]{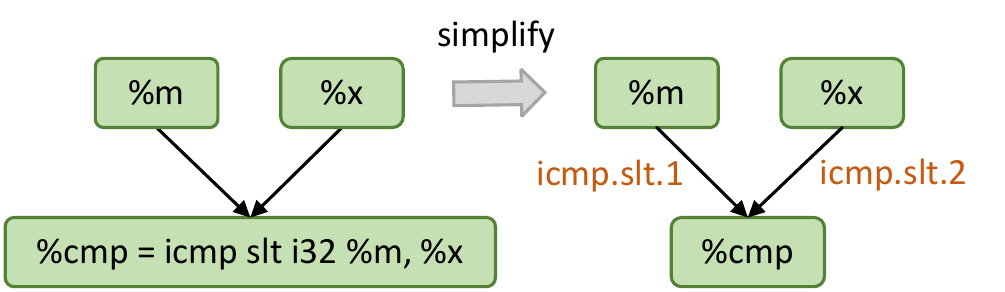}
    \label{figure:dfg_simplify}}
    \vfill
    \subfloat[Adding call graph to DFG.]
  {\includegraphics[width=\linewidth]{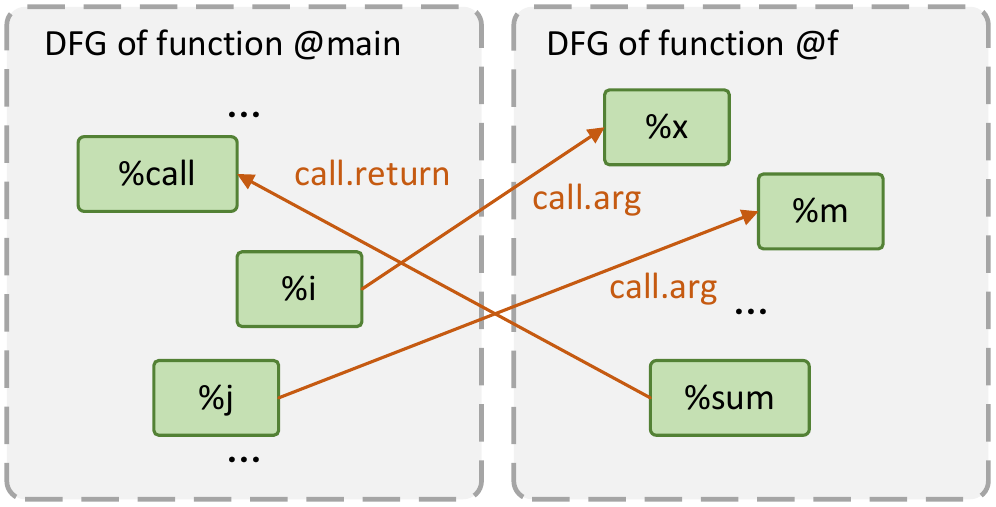}
  \label{figure:dfg_call}}
\end{minipage}
    \caption{The procedure of building input representation of the FAIR model.}
    \label{figure:input}
\end{figure*}

\section{FAIR}
\label{section:fair}

In this section, we present FAIR, including its \emph{Input} (Section~\ref{section:fair_input}), \emph{Architecture} (Section~\ref{section:fair_model}), and \emph{Pre-training Tasks} (Section~\ref{section:fair_tasks}).

\subsection{Input Representation}
\label{section:fair_input}

FAIR's input is constructed from a given IR program\footnote{Without loss of generality, we use LLVM IR (\url{https://llvm.org/docs/LangRef.html})}. Figure 1.a is an example of an IR function. Like most high-level programming languages, IR functions consist of a function signature and a function body, which contains one or more \textbf{Basic Blocks}, each starting with its label and a colon (e.g. ``entry:''). Each basic block consists of a sequence of \textbf{Instructions}, and the instructions in a basic block are executed sequentially, without any branches. 

Concretely, we propose a representation of an IR program that will be used as FAIR's input based on a Control Flow Graph (CFG) and a Data Flow Graph (DFG). This representation is composed of a \textit{CFG with Flow Type}, a \textit{Simplified DFG with Flow Type}, and \textit{BasicBlock-Variable Flows}.

The reason for encoding CFG and DFG separately instead of using CDFG is that CFG and DFG describe the behavior of the IR program from different perspectives, and they are completely independent of each other. Although CDFG is a graph formed by merging CFG and DFG, the information expressed by CFG and DFG is still independent and orthogonal in CDFG. Therefore, using one single neural network to encode two different types of information at the same time may lead to worse results. In addition, as a cost of merging, the CDFG becomes very large and there becomes more than one type of node, which makes it even more difficult for the neural network to encode it.

\subsubsection{CFG with Flow Type}
\label{section:fair_input_cfg}

A CFG specifies the order in which a function executes its instructions. It determines not only the sequence in which different parts of a function are executed but also how the function reacts to different conditions or inputs. The left side of Figure~\ref{figure:cfg} shows an original CFG of the IR function ``@main'' in Figure~\ref{figure:ir_example} (except ``br.T'' and ``br.F''). Each node of a CFG is a basic block and the edges are originally identical. The edges show possible jumps without the corresponding condition. For example, the basic block ``entry'' in this CFG has two jumps, <entry, if.end4> and <entry, for.body>\footnote{In this paper, we denote edges/flows without or with types as <$i$, $j$> or <$i$, $j$, $t$>, for untyped flows from node $i$ to $j$ or flows from $i$ to $j$ with type $t$, respectively.}. Using this CFG, it is impossible to determine which jump to choose. Therefore, we need to explicitly add this information to the CFG as the type of flow, which is important for understanding the jumps between basic blocks, and this is the reason why we consider CFGs to be heterogeneous graphs.

\textbf{\textit{Adding Flow Types.}} Such jump condition information can be retrieved from the last instruction in the basic block, i.e., the terminator instruction. Referring to Figure~\ref{figure:ir_example}, the terminator instruction of the basic block ``entry'' is a ``br'' instruction that is used to perform conditional or unconditional transfer between different basic blocks. This terminal instruction performs a conditional transformation using the previously computed Boolean variable ``\%cmp'' as the condition. If the condition is true, then it will jump to the basic block ``if.end4''; otherwise it will jump to ``for.body''. Based on this, we add the corresponding types, ``br.T'' and ``br.F'', to the two flows of the basic block ``entry'' in CFG, as shown in the orange words in Figure~\ref{figure:cfg}. It is worth mentioning that in addition to the ``br'' instruction, there are many other terminator instructions, such as ``ret'' (return to the caller), ``switch'', ``invoke'', etc.

\textbf{\textit{Adding Call Graphs.}} IR programs usually contain multiple functions, while the current CFG can only represent jump information inside a function. Therefore, we add call graph information between functions to associate them. The call graph shows which functions call which other functions and how they are connected. From Figure~\ref{figure:ir_example} in function ``@main'', one of the instructions of the basic block ``for.inc'' calls the function ``@f'', which executes the return instruction in its ``if.end'' basic block. In this case, we add two flows, <for.inc, @f-entry, call.func> and <if.end, for.inc, call.return>\footnote{We use ``@f-entry'' to avoid confusion with the basic block ``entry'' in ``@main''.}, indicating a function call and a return to the caller, respectively.

\subsubsection{Simplified DFG with Flow Type}
\label{section:fair_input_dfg}

DFG demonstrates the dependencies between instructions and values in a function. Figure~\ref{figure:dfg_simplify} shows how we add flow types to a DFG of the first instruction of the basic block ``entry'' of function ``@f'' in Figure~\ref{figure:ir_example}. On the left side, the original DFG consists of two types of nodes: variable/value nodes (e.g., ``\%m''), and instruction nodes. To unify the two types of nodes, we replace the entire instruction node with its return value, i.e., ``\%cmp''. By doing so, we can unify the nodes of the DFG into the same type, i.e., variables, which will make them easier to encode. Since this will lose information such as opcode, e.g. ``icmp slt'', we add opcode information as the flow types.

\textbf{\textit{Adding Flow Types.}} We assign the key information, i.e., opcode, which is discarded during the simplification of a DFG, to the type of flow. Concretely, we represent the opcode as three parts separated by dots: (1) \emph{opcodes}, such as ``icmp'' (integer comparison), ``add'', ``sub'', etc.; (2) \emph{options}, which are only available for certain operands, e.g. ``icmp'' has options such as ``eq'' (equal), ``ne'' (not equal), ``slt'' (signed less than), ``uge'' (unsigned greater or equal), etc., while ``add'', for example, has no options; and (3) \emph{operand positions}, which are only available for non-commutative operands/options, such as ``icmp.slt'', ``icmp.uge'', or ``sub'', but not for ``icmp.eq'', ``icmp.ne'', or ``add''. In this way, we complete the addition of DFG flow types that contain key information such as opcodes, options, and operand positions, namely <\%m, \%cmp, icmp.slt.1> and <\%x, \%cmp, icmp.slt.2> in the right of Figure~\ref{figure:dfg_simplify}.

\textbf{\textit{Add call graph.}} Just like a CFG, a DFG only represents the flow of data within one function. So, we add call graphs between different DFGs to join them together. Figure~\ref{figure:dfg_call} shows how to add the call graph. As can be seen in the example in Figure~\ref{figure:ir_example}, an instruction in the caller function ``@main'' calls the function ``@f'' with arguments ``\%i'' and ``\%j'', while the corresponding parameters of the callee is ``\%x'' and ``\%m''. Then the return instruction of the callee returns variable ``\%sum'', which is assigned to ``\%call'' in ``@main''. Therefore, we first add flows with type ``call.arg'' between the corresponding caller's arguments and callee's parameters, i.e., <\%i, \%x, call.arg> and <\%j, \%m, call.arg>. Then we add a flow from the callee's return variable to the caller's return value, with type ``call.return'', that is <\%sum, \%call, call.return>. Note that the flow type ``call.return'' here is not the same as the CFG flow type ``call.return''.

\subsubsection{BB-Var Flows}
\label{section:fair_input_bv}

Since we encode CFG and DFG separately, the relationship between the CFG and DFG is lost, and the relationship is most notably the subordination of Variables and Basic Blocks. This makes the construction of the BB-Var Flow (for connecting Basic Blocks to Variables) quite simple: if a variable node $m$ in the DFG belongs to one of the basic blocks $i$ in the CFG, then an untyped flow <$i$, $m$> will be added between the CFG and the DFG.

So far we have accomplished the processing and construction of the CFG, the DFG, and the BB-Var flow of an IR program. From now on we will consider the CFG and the DFG after the BB-Var flow connection as a whole graph $G=\langle V,F\rangle$, where $V$ is a set of CFG and DFG nodes, and $F=\{F_\textup{CFG},F_\textup{DFG},F_\textup{BV}\}$ is a set of all flows, where $F_\textup{CFG},F_\textup{DFG},F_\textup{BV}$ are sets of CFG flows, DFG flows and BB-Var flows, respectively.

\subsection{Model Architecture}
\label{section:fair_model}

\begin{figure}
    \centering
    \includegraphics[width=\linewidth]{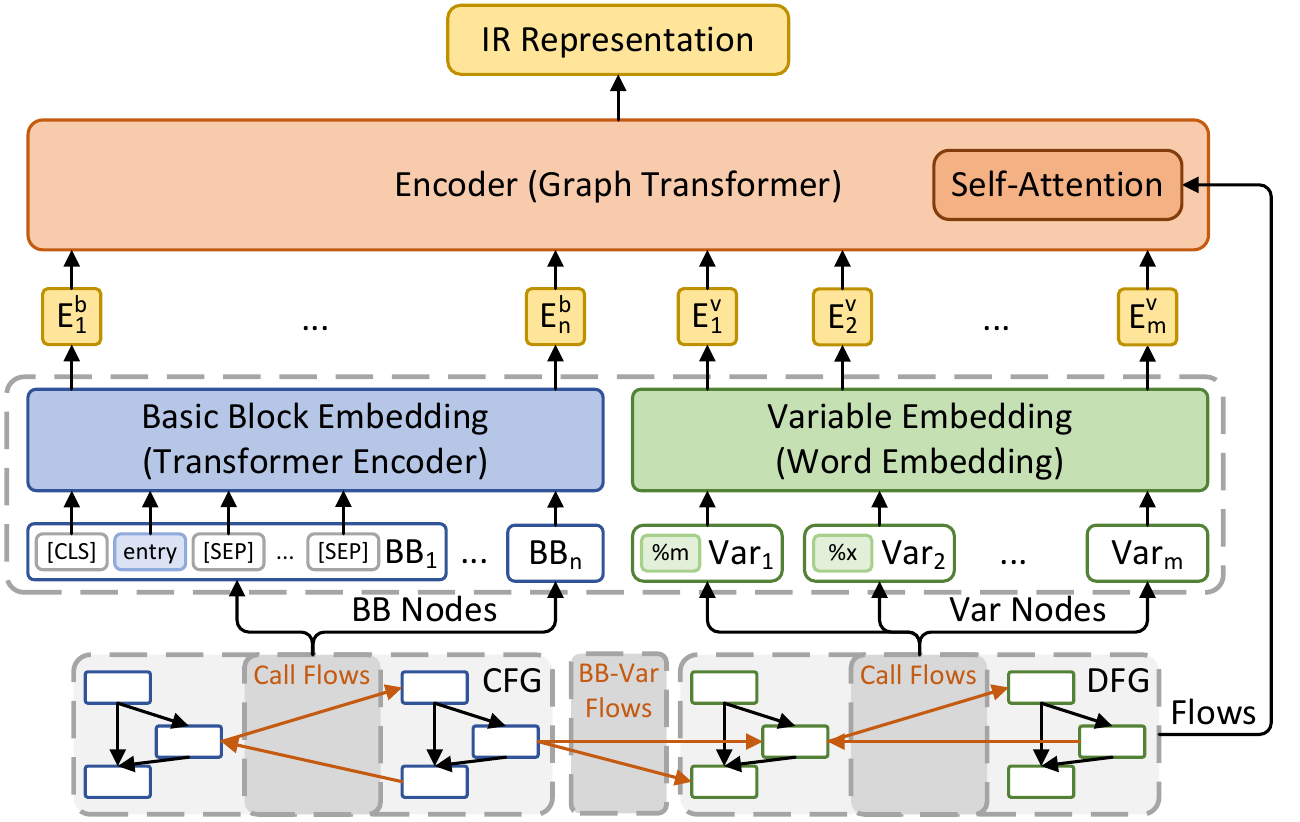}
    \caption{The overall architecture of FAIR model.}
    \label{figure:model}
\end{figure}

As shown in Figure~\ref{figure:model}, FAIR is a two-level model, where the first level, which includes Basic Block Embedding and Variable Embedding, is employed to encode the nodes in the CFG and the DFG to derive the embedding representation of the nodes, while the second level, the Encoder, is used to learn the overall IR representation from both the node embedding and the flow information within and between the CFG and the DFG.

\subsubsection{Node Embedding}
\label{section:fair_model_node}

The nodes of the CFG and the DFG are basic blocks and variables respectively, and given their distinct characteristics, we adopt different methods to embed these two kinds of nodes.

\textbf{\textit{Basic Block Embedding}}: 
As mentioned before, in basic blocks, instructions are executed sequentially, so we can naturally use a sequential model such as LSTM~\cite{hochreiter1997lstm} and Transformer Encoder~\cite{vaswani2017transformer} to encode a basic block. We choose a Transformer Encoder to embed the basic block in this paper.

We show in the bottom left of Figure~\ref{figure:model} how we obtain the embedding vector of each basic block. Given the CFG that we construct in Section~\ref{section:fair_input}, we first extract its nodes, namely the basic blocks, and present each basic block as a sequence of text tokens. Then, we add a special symbol ``[CLS]'' at the beginning of the sequence to identify the position of the output embedding vector, and feed this token sequence to the word embedding layer, the positional encoding layer, as well as several Transformer Encoder layers (in the figure, they are represented as a ``Basic Block Embedding''). Finally, we extract the hidden vector of ``[CLS]'' in the input at the last layer as the embedding vector of the whole basic block. Note that all basic blocks share the same Transformer Encoder when embedding.

In this manner, for a CFG, we can derive the embedding vectors of each node. These vectors are sorted in the order in which the basic blocks appear in the IR program, and the resulting sequence of vectors will be fed into the Encoder in the second level.

\textbf{\textit{Variable Embedding}}: 
The embedding of DFG nodes is relatively simple since these nodes are all variables. Specifically, we embed them using a regular Word Embedding layer. As shown in the bottom right of Figure~\ref{figure:model}, we (1) extract the variables from the processed DFG, (2) convert them into one-hot vectors, and (3) use a learnable linear layer to obtain the word embedding vectors of the variables.

\subsubsection{Encoder}
\label{section:fair_model_encoder}

We use Graph Transformer as the second level encoder to obviate the problems of long dependencies, over-smoothing, and over-squashing that are present in the message passing-based GNNs widely adopted in existing approaches~\cite{rampavsek2022recipe,min2022transformer,kim2022pure}. The two inputs of this encoder are the output of the first-level encoder and the flow information. Note that they are utilized in different ways.  

\textbf{\textit{Formulate Node Embedding}}: 
Given a sequence of $m$ vectors of basic blocks $E^b=[E^b_1,\dots,E^b_m]\in \mathbb{R}^{m\times d}$, and a sequence of $n$ vectors of variables $E^v=[E^v_1,\dots,E^v_n]\in\mathbb{R}^{n\times d}$, where $d$ denotes the hidden dimension of our model. We first build the input of the Encoder. Specifically, we concatenate these two sequences,  using the embedding vector of ``[SEP]'' ($E_{\textup{[SEP]}}\in\mathbb{R}^d$) to separate them, and insert the embedding vectors of ``[CLS]'' ($E_{\textup{[CLS]}}\in\mathbb{R}^d$) and $E_{\textup{[SEP]}}$ at the beginning and the end of the sequence, respectively. Consequently, we form the input to the Encoder $I_\textup{Enc}\in\mathbb{R}^{l\times d}$, where $l=m+n+3$,

\begin{equation}
    I_\textup{Enc}=[E_{[\textup{CLS}]},E^b_1,\dots,E^b_m,E_{\textup{[SEP]}},E^v_1,\dots,E^v_n,E_{\textup{[SEP]}}]
\end{equation}

In order for the Encoder to better distinguish the nodes of the CFG and those of the DFG, we add another vector sequence $I_\textup{type}\in\mathbb{R}^{l\times d}$ on top of $I_\textup{Enc}$ before we input this vector sequence to the Encoder. This is achieved by a mechanism similar to Segment Embedding in BERT~\cite{devlin2019bert}. To be specific, we construct a sequence containing only 0s and 1s, where a 0 is used to indicate a CFG node (i.e., the position of $E^b$ in the $I_{Enc}$ (including the special symbols)), while a 1 is used to indicate a DFG node (i.e., the position where $E^v$ is located). Then we pass the numbers in this sequence through another embedding layer and get the vector sequence that presents the node type, i.e. $I_{type}$.

Finally, we add $I_\textup{Enc}$ and $I_\textup{type}$ and input the results $I=I_\textup{Enc}+I_\textup{type}\in\mathbb{R}^{l\times d}$ into the Encoder.

\textbf{\textit{Integrating Flows}}: 
We make the model learn flow information by injecting graph priors into the attention computation via graph bias terms. In other words, since our input is composed of the nodes of the graph, when we compute the self-attention matrix in each layer of the Transformer, the flow information between the nodes is injected into the attention matrix through an adjacency matrix. This makes our model different from the vanilla Transformer Encoder~\cite{vaswani2017transformer} in the self-attention module of each encoder layer. For simpler illustration without loss of generality, we assume in this section that there is only one self-attention head.

Concretely, let $H=[h_1,\dots,h_l]\in \mathbb{R}^{l\times d}$ be the input of the self-attention module, where $h_i\in\mathbb{R}^d$ is the hidden vectors of position $i$. The attention scores of input matrix $H$ are computed as:
\begin{align}
    Q=HW_Q,~K=HW_K,~V=HW_V,
    \\
    \textup{Attention}(H)=\textup{softmax}(A)V,
\end{align}
where $W_Q,W_K,W_V\in\mathbb{R}^{d\times d}$ are projection matrices, $A\in\mathbb{R}^{l\times l}$ is the attention score matrix between every two input nodes. Let $A_{ij}$ be the (i, j) elements of $A$, we have:
\begin{align}
    \label{equation:attention_score}
    A_{ij}&=\frac{(h_iW_Q)(h_jW_K)^T}{\sqrt{d}}+b,\\
    b&=\begin{cases}
    0, & \langle i,j\rangle\notin F \\
    b^\textup{CFG}_{\phi(t)}, & \langle i,j,t\rangle\in F_\textup{CFG} \\
    b^\textup{DFG}_{\phi(t)}, & \langle i,j,t\rangle\in F_\textup{DFG} \\
    b^{BV}, & \langle i,j\rangle\in F_\textup{BV} \\
    \end{cases},
\end{align}
where $b^\textup{CFG}_{\phi(t)},b^\textup{DFG}_{\phi(t)}\in\mathbb{R}$ are learnable parameters indexed by $\phi(t)$. Taking a CFG as an example, we let there be a total of $p$ CFG flow types. Then we have a vector $B^\textup{CFG}=[b^\textup{CFG}_1,\dots,b^\textup{CFG}_p]\in\mathbb{R}^p$, and $\phi(t)$ is the index of CFG flow type $t$ in $B^\textup{CFG}$. The vector $b^{BV}\in\mathbb{R}$ is also learnable. All these three types of parameters are shared in all layers. It can be seen that we achieve the injection of flow information by adding bias terms to the attention scores. Specifically, when calculating the attention score between nodes $i$ and $j$, if there is no flow between them, we do not add a bias term, but if there is a flow of CFG or DFG of type $t$ between them, then we add the bias term corresponding to that type $t$ to the attention score, noting that there is a corresponding learnable bias term for each flow type of CFG and DFG. Finally, if there is an untyped flow of BB-Var between $i$ and $j$, we add another learnable bias term to the attention score.

With respect to the other aspects, e.g., the feed-forward module and layer normalization, FAIR is identical to the vanilla Transformer Encoder~\cite{vaswani2017transformer}, so we will not go over them here. Next, we present the pre-training tasks used to train FAIR.

\subsection{Pre-Training Tasks}
\label{section:fair_tasks}

Pre-training has been shown to massively improve the performance of models on downstream tasks~\cite{devlin2019bert,liu2019roberta,feng2020codebert,niu2022sptcode}. With respect to Graph Transformer, pre-training is able to help a model to learn generalizable and transferable representations of graphs and exploit additional knowledge to guide a model to capture structural and semantic information~\cite{li2022kpgt,liu2022mask}. Therefore, we propose five pre-training tasks that enable the model to learn the semantic information in the basic block, the flow information of the graph, and the overall representation capability for IR.

\subsubsection{Masked Language Modeling}
\label{section:fair_tasks_mlm}

Masked Language Modeling (MLM) is widely adopted in the field of NLP and SE~\cite{devlin2019bert,clark2019electra,feng2020codebert,guo2022unixcoder}. It can help a model to acquire a good contextual and semantic understanding of the basic block~\cite{liu2019roberta,van2021masked}. As a result, we first adopt MLM to train our Basic Block Embedding module to generate better representations for basic blocks. The task is to predict the original tokens that are masked in the input. We follow the original MLM setup, which samples 15\% of the tokens from the input sequence, and then replaces 80\% of them with a [MASK] token, 10\% with a random token, and another 10\% of them unchanged.

Let $x=[x_1,\dots,x_n]$ be a sequence of tokens of a basic block of length $n$ and $M$ be a set of indices of masked tokens. Then the MLM objective is to minimize the following loss:
\begin{equation}
    \mathcal{L}_\textup{MLM}=-\frac{1}{|M|}\sum_{i\in M}\log P(x_i|x_{\neg i}),
\end{equation}
where $x_{\neg i}$ denotes the sentence $x$ with the $i$-th token masked and $P(x_i|x_{\neg i})$ denotes the probability of predicting the original token $x_i$ given the masked sentence $x_{\neg i}$.

\subsubsection{CFG/DFG Flow Type Prediction}
\label{section:fair_tasks_cft_dft}

To learn the added flow information in a CFG and a DFG, we design two pre-training tasks, one for each of these two graphs, namely CFG Flow Type Prediction (CFT) and DFG Flow Type Prediction (DFT). We adopt these two pre-training tasks with the motivation that the model learns the structure-aware information of the input IR so that it can grasp the control flow information in the CFG and the data flow information in the DFG. The objectives of these two tasks are the same, so we only illustrate CFT here for the sake of simplicity.

Given a CFG with $n$ nodes, we randomly sample 15\% from a total of $n^2$ ordered pairs of nodes, i.e., $n_m=15\%\times n^2$. Then, we mask the flows if they exist between these pairs, and subsequently, we make the model predict whether these edges exist as well as the type of edges. The absence of a flow can be seen as a special type of flow, and consequently, this task becomes a $(k+1)$-way classification task, where the number of CFG flow types is $k$. Note that each type of flow is sampled in a balanced manner.

Formally, let $F^\textup{CFG}_m=\{\langle u_i,v_i,\phi_i\rangle |i\in[1,n_m],\phi_i\in [0,k]\}$ be the set of sampled node pairs, where $\phi_i$ indicates the index of the flow type (with 0 representing no flow and [1,$k$] representing the flow types). Therefore, the masked CFG becomes $G^\textup{CFG}_m=\langle V^\textup{CFG},F^\textup{CFG}\setminus F^\textup{CFG}_m\rangle$. where $F^\textup{CFG}\setminus F^\textup{CFG}_m$ represents the set difference between the original set of flows $F^\textup{CFG}$ and the masked flows $F^\textup{CFG}_m$. 

Assuming $\langle u_i,v_i,\phi(t)\rangle$ is the $i$-th element in $F^\textup{CFG}_m$, the model predicts the type of flow by inputting their hidden vectors in the last layer into an extra linear layer. Let $h_{u_i},h_{v_i}\in\mathbb{R}^d$ be the hidden vectors of nodes $i$ and $j$. The index of the predicted flow type $\hat{\phi_i}\in[0,k]$ is:
\begin{equation}
    \hat{\phi_i}=\argmax (\textup{softmax}(W[h_{u_i};h_{v_i}]+b)),
\end{equation}
where $W\in\mathbb{R}^{2d\times (k+1)}$ and $b\in\mathbb{R}^{k+1}$ are learnable parameters, and $[h_{u_i};h_{v_i}]$ is the concatenation of $h_{u_i}$ and $h_{v_i}$. We can then represent the predicted index set of masked flows as $\hat{F^\textup{CFG}_m}=\{\phi_i|i\in[1,|E_m|]\}$. The objective of CFT is to minimize the following loss:
\begin{align}
    \mathcal{L}_\textup{CFT}=-\frac{1}{|F^\textup{CFG}_m|}\sum_{\langle u,v,\phi\rangle\in F^\textup{CFG}_m}\log P(\phi|G^\textup{CFG}_m),
\end{align}
where $P(\phi|G_m)$ is the probability of predicting the original flow type $\phi$ given the masked CFG $G_m$.

In the same way, the DFT objective is to minimize the following loss:
\begin{align}
    \mathcal{L}_\textup{DFT}=-\frac{1}{|F^\textup{DFG}_m|}\sum_{\langle u,v,\phi\rangle\in F^\textup{DFG}_m}\log P(\phi|G^\textup{DFG}_m),
\end{align}
where $F^\textup{DFG}_m$ is the set of sampled node pairs, $G^\textup{DFG}_m$ denotes the DFG after masking the flow type in $F^\textup{DFG}_m$.

\subsubsection{BB-Var Flow Prediction}
\label{section:fair_tasks_bvp}

BB-Var Flow Prediction (BVP) is similar to CFT and DFT, except that BVP is a binary classification task that only predicts whether the flow exists or not. Let $G^\textup{BV}=\langle V^\textup{BV},F^\textup{BV}\rangle$ denote the graph where $V^\textup{BV}$ includes $n$ basic blocks and $m$ variable nodes, and $F^\textup{BV}$ is the set of BB-Var flows. We use the same probability (i.e., 15\%) to mask the flow in $G^\textup{BV}$, which results in the masked graph $G^\textup{BV}_m=\langle V^\textup{BV},F^\textup{BV}\setminus F^\textup{BV}_m$, where $F^\textup{BV}_m$ is the set of masked flows. The loss of BVP is calculated as:
\begin{gather}
    p_{\langle u,v\rangle}=\textup{sigmoid}(h_u\cdot h_v^T),\\
    \mathcal{L}_\textup{BVP}=-\frac{1}{|F^\textup{BV}_m|}\sum_{\langle u,v\rangle\in F^\textup{BV}_m}[y\log p_{\langle u,v\rangle}+(1-y)\log (1-p_{\langle u,v\rangle})],
\end{gather}
where $h_u,h_v\in\mathbb{R}^d$ are the hidden vectors of the nodes $u$ and $v$ in the last layer, $y$ is 1 if $\langle u,v\rangle\in F^\textup{BV}$ and 0 otherwise, and $p_{\langle u,v\rangle}$ is the probability of predicting that there is an BB-Var flow between nodes $u$ and $v$.

\subsubsection{Contrastive Learning}
\label{section:fair_tasks_cl}

We employ contrastive learning as our last pre-training task. Contrastive learning aims to learn representations of an input example (a.k.a. the anchor example) by contrasting its positive and negative pairs\footnote{Positive examples are examples that are similar to the anchor example, while negative examples are examples that are different from the anchor examples. The goal is to make the positive examples closer to the anchor example and the negative examples farther away in the representation space.}, which allows models to improve their capabilities on multiple dimensions, such as scalability~\cite{bojchevski2020scaling}, generalization ability~\cite{you2020graph}, global and hierarchical local features learning~\cite{chen2021intriguing} and performance on downstream tasks~\cite{chen2020big,zhang2022contrastive}.

The key to contrastive learning is to construct positive and negative examples. Recall that the input of our model can be seen as a single graph (Section~\ref{section:fair_input_bv}) $G$. Given $G$, we leverage the following methods to construct positive examples.

\begin{itemize}
    \item \textbf{Function Permutation}: randomly change the order of the functions when input contains multiple functions (which is the majority of cases).
    \item \textbf{Function down-sampling}: remove one or more functions randomly when the input contains more than one function.
    \item \textbf{Flow Mutation}: randomly change some of the flows, i.e. adding, removing flows, and changing the type of the flows.
    \item \textbf{Node Adding/Removing}: add some random standalone nodes (no flow), or remove some nodes (and its flows).
\end{itemize}

Since most of our inputs contain multiple functions, it is natural to construct positive examples by treating each function as a subgraph, such as function down-sampling, rather than the random down-sampling used by other methods.

Unlike OSCAR and IRGen, which also use contrastive learning based on IR, FAIR constructs positive examples using the aforementioned methods instead of using different optimization options to generate different IRs from the same source code. We believe that the method of constructing positive examples using different optimization options would lead to exposure bias of the model on a downstream task, i.e., the model is only trained on IRs generated by different optimization options during pre-training, while on downstream tasks, IRs are generated after the same optimization options. This could result in a model that only learns how to identify IRs generated by the same source through different optimization options, rather than different IRs generated by different sources through the same optimization options.

Some might also argue that when we construct positive examples by using the methods described above, the underlying IR of constructed positive examples could be incorrect and semantically invalid. While this may be true, we think that it does not impact the effectiveness of the contrastive learning we use. Our goal with contrastive learning is to enable the model to distinguish similar and dissimilar examples, and semantically similar IRs will then result in similar graph representations. As the input to our model is the graph, we, therefore, are able to make some changes to the graph of the anchoring example to construct pseudo-graphs with similar graph structures, without requiring a large-scale dataset with real-world similar IRs.

As negative examples, we utilize other examples in the training mini-batch. Then, we feed the input and the positive/negative examples into the model and obtain their representations. Let $v\in\mathbb{R}^d$ denote the representation vector of the $G$, and $S_\textup{pos}=\{v^\textup{pos}_1,\dots,v^\textup{pos}_n\}$, $S_\textup{neg}=\{v^\textup{neg}_1,\dots,v^\textup{neg}_m\}$ be the sets of representations of positive and negative examples, respectively. The loss of contrastive learning is computed as follows,
\begin{equation}
    \mathcal{L}_\textup{CL}=\textup{max}(0, D_\textup{pos} - D_\textup{neg} + \textup{margin}),
\end{equation}
where $D_\textup{pos},D_\textup{neg}\in\mathbb{R}$ are the averages of the Euclidean distance~\cite{dokmanic2015euclidean} between $v$ and each element in $S_\textup{pos}$ and $S_\textup{neg}$, respectively.

\subsubsection{Overall Objective}

The overall pre-training objective is to minimize the sum of the all above losses, that is,
\begin{equation}
    \mathcal{L}=\mathcal{L}_\textup{MLM}+\mathcal{L}_\textup{CFT}+\mathcal{L}_\textup{DFT}+\mathcal{L}_\textup{BVP}+\mathcal{L}_\textup{CL}
\end{equation}

\section{Evaluation Setup}
\label{section:exp}

\subsection{Pre-Training}
\label{section:exp_pretrain}

\textit{\textbf{Data Preparation.}}
\label{section:exp_pretrain_dataset}
We adopt the dataset provided by Peng et al.~\cite{peng2021oscar} as our pre-training dataset, which consists of eleven popular open-source C/C++ projects from GitHub. This dataset includes 41,322 IR programs, 855,792 functions, and 48,023,781 instructions in total. We further optimize the given IRs using LLVM of version 13.0.1 with the optimization options ``-Os'' and ``-ffast-math''.

\textit{\textbf{Tokenizer.}}
\label{section:exp_pretrain_settings}
Due to the large gap between the lexical features of IR and those of the high-level programming languages, we do not use existing tokenizers developed for high-level languages. Instead, we build a tokenizer of size 30,000 from scratch using the BPE algorithm~\cite{sennrich2016bpe} upon the pre-training data.

\textit{\textbf{Hyperparameters.}}
We set the hidden dimension $d$ to 768, the intermediate dimension of feed-forward to 3072, the number of layers of the Basic Block Embedding module and Encoder module to 6, and the number of self-attention heads to 12. We set the maximum length of each basic block to 256, the maximum number of basic blocks of each program (which is also the number of CFG nodes) to 64, and the maximum number of DFG nodes to 256. This results in a total of 138M parameters used for model pre-training, of which 30M are temporary parameters that are only used during pre-training. This gives us 108M pre-trained model parameters for the downstream tasks. We pre-train FAIR for 10 epochs by minimizing the loss $\mathcal{L}$. We use AdamW~\cite{loshchilov2018adamw} as our optimizer. The initial learning rate is 5e-5 and the warmup step is 2,000. The pre-training is run on 4 NVIDIA V100 32G GPUs with a total batch size of 8.

\subsection{Downstream Tasks}
\label{section:exp_downstream}

In this subsection, we present the fine-tuning procedure of FAIR on four downstream tasks. For each downstream task, we first provide a brief introduction and then describe the dataset and the evaluation metrics.

\subsubsection{Code-to-Code (C2C) Retrieval}
\label{section:exp_downstream_cr}

Given a source code as the query, the code-to-code retrieval task aims to retrieve codes with the same semantics from a collection of candidates. This task can evaluate the ability of a model to distinguish between codes/IRs with different semantics. 

We use two datasets for this task, namely, POJ-104~\cite{mou2016poj104} and GCJ\footnote{https://github.com/Jur1cek/gcj-dataset}. POJ-104 contains 42,000 C/C++ programs that implement entry-level programming assignments for 104 different problems. We use the train/valid/test splits provided by CodeXGLUE~\cite{lu2021codexglue}, where the numbers of problems/codes of each split are 64/32,000, 16/8,000, and 24/12,000. GCJ contains the source code from solutions to Google Code Jam programming challenges and includes 302,070 C/C++ programs across 331 problems. There are no available splits, so we create the train/valid/test splits, which include 265/26/40 problems and 181,103/60,230/60,737 programs.

As for the metric, we adopt mean average precision with the recall level of 499 (i.e., MAP@R, R=499)~\cite{lu2021codexglue}. That is, we let the model retrieve the top 499 semantically similar candidates given a query.

\subsubsection{Algorithm Classification}
\label{section:exp_downstream_class}

Algorithm classification aims to categorize a given code. We also use the POJ-104 as the dataset, but adopt the train/valid/test split created by Ben et al.~\cite{ben2018ncc}. The sizes of train/valid/test splits are 27,649/9,155/9,227. We use the error rate (ER) on the test set as the evaluation metric.

\subsubsection{Heterogeneous Device Mapping}
\label{section:exp_downstream_device}

Heterogeneous device mapping is the task of choosing the execution device that has the best performance given an \textit{OpenCL Kernel}, the \textit{Input Data Size} and \textit{Work Group Size} (i.e., the number of threads that work in a group with shared memory). We use the dataset provided by Grewe et al.~\cite{grewe2013portable}, who formulate this task as a binary classification task. This dataset consists of two subtasks, namely predicting whether the given OpenCL kernel will run faster on an Intel CPU or an AMD GPU and whether it will run faster on an Intel CPU or an NVIDIA GPU. Both of them contain 680 labeled examples derived from the 256 unique kernels by varying dynamic inputs.

In addition to accuracy (Acc), we use a metric called ``Speedup'', which is the average ratio of the runtime improvement of each OpenCL on the devices predicted by the model compared to the runtime of the static mapping. The static mapping chooses CPU when comparing CPU and AMD GPU, and chooses GPU when comparing CPU and NVIDIA GPU.

We concatenate the \textit{Input Data Size} and \textit{Work Group Size} to create the input. Following the usual strategy of utilizing this dataset~\cite{grewe2013portable,ben2018ncc,cummins2021programl}, we use 10-fold cross-validation with rotating 8/1/1 train/valid/test splits for evaluation.

\subsubsection{Optimal Thread Coarsening Factor}
\label{section:exp_downstream_thread}

Given an OpenCL kernel, this task is to predict the best-performing thread coarsening factor, which is a value that determines how many threads to merge together. 

We adopt the dataset provided by~\cite{magni2014automatic}. It contains the runtimes on 17 benchmarks with 4 GPUs having thread coarsening factors of 1, 2, 4, 8, 16, and 32, respectively. The GPUs are Cypress (AMD Radeon HD 5900), Tahiti (AMD Tahiti 7970), Fermi (NVIDIA GTX 480) and Kepler (NVIDIA Tesla K20c). It is a 6-way classification task (i.e., predicting one of the 6 possible factors) and includes 4 subtasks, each corresponding to one GPU.

We use the Speedup metric to evaluate the performance of the model. Speedup is the ratio of runtime reduction of the GPU at the factor predicted by the model to the runtime without thread coarsening (i.e., when the factor is 1).

\subsection{Fine-Tuning}
\label{section:exp_finetune}

The pre-trained FAIR model will be fine-tuned on each individual downstream task. We discard the modules that are temporarily added during the pre-training phase, such as the learnable matrix $W$ and the vector $b$ in the classification head module (see Section~\ref{section:fair_tasks_cft_dft}), and only preserve all the modules present in Figure~\ref{figure:model} when FAIR is applied to the downstream tasks. For the classification model, we will add the corresponding classification module so that the representation vector generated by FAIR can be mapped to each class. Before fine-tuning, we convert high-level source code into LLVM IR for each dataset of the downstream tasks with Clang 13.0.1. LLVM 13.0.1 is used to optimize the LLVM IR.

\subsection{Baselines}
\label{section:exp_baselines}

We use two groups of baselines. The first group is composed of models of high-level language source code, all of which were pre-trained on source code and have achieved state-of-the-art performance on various code-related downstream tasks. They are \textbf{CodeBERT}~\cite{feng2020codebert}, \textbf{CodeT5}~\cite{wang2021codet5} and \textbf{UniXcoder}~\cite{guo2022unixcoder}. For each downstream task, these three models are directly fine-tuned on the high-level source code in the dataset. The second group is composed of models that are designed for IR, including \textbf{ncc}, \textbf{IR2VEC}, \textbf{GNN-CDFG}, \textbf{ProGraML}, \textbf{OSCAR}, and \textbf{IRGen}. They are introduced in Section~\ref{section:related_ir}.

\section{Results and Discussion}
\label{section:exp_eval}

To evaluate FAIR, we propose three Research Questions. We run each experiment three times by using different random seeds and report the mean. To check the statistical significance of the experimental results, we utilize the Approximate Randomization Test\footnote{\url{https://github.com/danieldk/approx-rand-test}}.

\subsection{Comparison with Baselines}
\label{section:exp_eval_rq1}

\textbf{RQ1: How effective is FAIR compared with the state-of-the-art baselines on four downstream tasks?}

We conduct experiments to check the performances of all compared approaches on the four downstream tasks. The results of code-to-code retrieval and algorithm classification are in Table~\ref{table:results_cr_cls}, and the results of heterogeneous device mapping and optimal thread coarsening factor are in Tables~\ref{table:results_device} and~\ref{table:results_thread}, respectively. (Note that in these tables, (1) the best results are boldfaced, and (2) the differences between the best result and the other results are statistically significant at $p < 0.05$.) Overall, FAIR achieves either new SOTA performance or performance comparable to the current SOTA models on all four downstream tasks.

\begin{table}[t]
    \centering
    \caption{Results on C2C retrieval and algorithm classification.}
    \label{table:results_cr_cls}
    \begin{tabular}{lrrr}
        \toprule
            \multirow{3}{*}{Models} &
            \multicolumn{2}{c}{Retrieval} &
            \multicolumn{1}{c}{\multirow{2}{*}{\begin{tabular}[c]{@{}c@{}}Algorithm\\ Classification\end{tabular}}} \\
        \cline{2-3}
            &
            \multicolumn{1}{c}{POJ-104} &
            \multicolumn{1}{c}{GCJ} &
            \multicolumn{1}{c}{} \\
        \cline{2-4}
            & MAP@R          & MAP@R          & Error Rate    \\
        \midrule
            CodeBERT  & 82.67                       & 77.16                     & 4.61                           \\
            CodeT5    & 88.65                       & 79.65                     & 4.12                           \\
            UniXcoder & 90.52                       & 82.23                     & 1.91                           \\
        \hline
            ncc       & 54.19                       & 64.68                     & 5.17                           \\
            IR2Vec    & 76.34                       & 77.90                      & 3.93                           \\
            GNN-CDFG  & 79.20                       & 66.64                     & 3.72                           \\
            ProGraML  & 81.53                       & 71.27                     & 3.38                           \\
            OSCAR     & 89.98                       & 81.76                     & 1.92                           \\
            IRGen     & 89.22                       & 83.26                     & 2.01                           \\
        \hline
            FAIR      & \textbf{92.04}              & \textbf{85.41}            & \textbf{1.75}                  \\
        \bottomrule
    \end{tabular}
\end{table}
\begin{table}[t]
    \centering
    \caption{Results on heterogeneous device mapping.}
    \label{table:results_device}
    \begin{tabular}{lrrrr}
        \toprule
            \multirow{2}{*}{Models} & \multicolumn{2}{c}{NVIDIA}        & \multicolumn{2}{c}{AMD}     \\
        \cline{2-5}
                                    & Acc            & Speedup       & Acc            & Speedup       \\
        \midrule
            CodeBERT                & 86.76          & 1.58          & 95.59          & 2.79          \\
            CodeT5                  & 88.54          & 1.48          & 93.10          & 2.59          \\
            UnixCoder               & 89.71          & 1.50          & 94.12          & 2.76          \\
        \hline
            ncc                     & 84.67          & 1.44          & 88.09          & 3.47          \\
            IR2Vec                  & 85.32          & 1.26          & 91.32          & 3.51          \\
            GNN-CDFG                & 87.93          & 1.39          & 89.16          & 3.37          \\
            ProGraML                & 88.13          & 1.41          & 92.60          & 2.98          \\
            OSCAR                   & 89.52          & 1.49          & 94.11          & 3.34          \\
            IRGen                   & 89.86          & 1.57          & 94.32          & 3.60          \\
        \hline
            FAIR                    & \textbf{91.61} & \textbf{1.62} & \textbf{96.52} & \textbf{3.63} \\
        \bottomrule
    \end{tabular}
\end{table}
\begin{table}[t]
    \centering
    \caption{Results on optimal thread coarsening factor. }%Models in the first group are omitted, because they always predict the same labels.
    \label{table:results_thread}
    \begin{tabular}{lrrrr}
        \toprule
            Models    & Cypress       & Tahiti        & Fermi         & Kepler        \\
        \midrule
            ncc       & 1.01          & 1.04          & 0.95          & 1.01          \\
            IR2Vec    & 1.18          & \textbf{1.21} & 1.1           & \textbf{1.08} \\
            GNN-CDFG  & 1.01          & 0.93          & 0.92          & 0.86          \\
            ProGraML  & 1.05          & 1.12          & 0.96          & 0.97          \\
            OSCAR     & 1.21          & 1.19          & 1.06          & 1.07          \\
            IRGen     & 1.22          & 1.17          & 1.11          & \textbf{1.08} \\
        \hline
            FAIR      & \textbf{1.25} & \textbf{1.21} & \textbf{1.13} & \textbf{1.08} \\
        \bottomrule
    \end{tabular}
\end{table}
\begin{table*}[t]
    \centering
    \caption{Ablation results on downstream tasks. The best results in each column are boldfaced, and the worst results in each group are underlined. In cases where the model predicts the same label for all examples, the result is replaced with a '-'.}
    \label{table:results_ablation}
    \resizebox{\linewidth}{!}{%
    \begin{tabular}{lrrrrrrrrrrr}
        \toprule
            \multirow{3}{*}{Methods} &
            \multicolumn{2}{c}{Retrieval} &
            \multicolumn{1}{c}{\multirow{2}{*}{\begin{tabular}[c]{@{}c@{}}Algorithm\\ Classification\end{tabular}}} &
            \multicolumn{4}{c}{Device Mapping} &
            \multicolumn{4}{c}{Thread Coarsening Factor} \\
        \cline{2-3}\cline{5-12}
            &
            \multicolumn{1}{c}{POJ-104} &
            \multicolumn{1}{c}{GCJ} &
            \multicolumn{1}{c}{} &
            \multicolumn{2}{c}{AMD} &
            \multicolumn{2}{c}{NVIDIA} &
            \multicolumn{1}{c}{Cypress} &
            \multicolumn{1}{c}{Tahiti} &
            \multicolumn{1}{c}{Fermi} &
            \multicolumn{1}{c}{Kepler} \\
        \cline{2-12}
            &
            MAP@R &
            MAP@R &
            ER &
            Acc &
            Speedup &
            Acc &
            Speedup &
            Speedup &
            Speedup &
            Speedup &
            Speedup \\
        \midrule
            FAIR &
            {\bf 92.04} &
            {\bf 85.41} &
            {\bf 1.75} &
            {\bf 91.61} &
            {\bf 1.62} &
            {\bf 95.52} &
            3.63 &
            {\bf 1.13} &
            {\bf 1.25} &
            {\bf 1.21} &
            {\bf 1.08} \\
        \hline
            ~-w/o type &
            90.32 &
            83.50 &
            1.79 &
            91.23 &
            1.59 &
            95.11 &
            3.61 &
            {\bf 1.13} &
            1.24 &
            {\bf 1.21} &
            {\bf 1.08} \\
        ~-w/o flow &
            88.95 &
            81.83 &
            1.94 &
            \underline{90.66} &
            \underline{1.48} &
            \underline{94.43} &
            \underline{3.47} &
            \underline{1.12} &
            \underline{1.22} &
            \underline{1.20} &
            {\bf 1.08} \\
        ~-w/ CDFG &
            \underline{87.13} &
            \underline{79.39} &
            \underline{2.48} &
            91.01 &
            1.56 &
            95.09 &
            3.58 &
            {\bf 1.13} &
            1.24 &
            \underline{1.20} &
            \underline{1.07} \\
        \hline
            ~-w/o MLM &
            91.85 &
            84.94 &
            1.91 &
            \underline{89.02} &
            \underline{1.40} &
            \underline{93.85} &
            \underline{3.25} &
            - &
            1.21 &
            - &
            - \\
        ~-w/o xFT &
            91.14 &
            84.86 &
            2.03 &
            91.38 &
            1.56 &
            95.16 &
            3.34 &
            \underline{1.11} &
            \underline{1.18} &
            1.19 &
            {\bf 1.08} \\
        ~-w/o BVP &
            91.25 &
            85.19 &
            1.99 &
            91.6 &
            1.59 &
            {\bf 95.52} &
            {\bf 3.64} &
            1.12 &
            1.21 &
            {\bf 1.21} &
            {\bf 1.08} \\
        ~-w/o CL &
            \underline{88.09} &
            \underline{81.61} &
            \underline{2.88} &
            90.96 &
            1.58 &
            95.15 &
            3.59 &
            \underline{1.11} &
            \underline{1.18} &
            \underline{1.18} &
            \underline{1.06} \\
        \hline
            ~-w/o all &
            87.36 &
            79.14 &
            2.98 &
            - &
            - &
            - &
            - &
            - &
            - &
            - &
            - \\
        \bottomrule
    \end{tabular}%
    }
\end{table*}

In addition to that FAIR achieves new SOTA for the code-to-code retrieval task, Table~\ref{table:results_cr_cls} also shows that pre-trained models of both source code (i.e., CodeBERT, CodeT5, and UnixCoder) and IR (i.e., OSCAR, IRGen, and FAIR) generally achieve higher performance then non-pre-training approaches (i.e., ncc, IR2Vec, GNN-CDFG, and ProGraML) on both of the two datasets. Comparing the performance of each approach on different datasets, we find that the ncc and IR2Vec that use lookup tables perform better on GCJ than on POJ-104, while the others perform better on POJ-104 than on GCJ.

Examining the results in Table~\ref{table:results_device}, we find that the pre-trained models (i.e., the first group of models, OSCAR, IRGen, and FAIR) tend to have better performance than their non-pre-trained counterparts. Since the dataset for this task is small (with only 680 examples per subtask), we speculate that pre-training can help a model learn more general features and more transferable representations from large-scale data and subsequently improve its performance on a downstream task that has insufficient data~\cite{abnar2021exploring,liu2021multi,cui2022contrastive}.

However, a much smaller amount of data occurs in the optimal thread coarsening factor task in Figure~\ref{table:results_thread}, where each subtask has only 17 examples. We find that the IR-based pre-trained model continues to have better performance than the others. Note that even a model as small and shallow as IR2VEC has remarkable performance, possibly because small models require fewer data to train and are also less likely to overfit the training data~\cite{ba2014deep,zhang2017understanding}. We do not show the results of the models of the first group in Table~\ref{table:results_thread} because they always predict the same label for all examples. One reason for this behavior is that the task is a multi-label classification task, which has a large gap with the pre-training tasks used to pre-train these models. Another reason is the data distribution gap: these models are all pre-trained on CodeSearchNet~\cite{husain2019codesearchnet} (or plus C/C\# from BigQuery~\cite{wang2021codet5}), which does not contain OpenCL kernel-related code. Above all, having too little data prevents them from effectively transferring the code representation to this task.

\subsection{Model Ablation}
\label{section:exp_eval_rq2}
\textbf{RQ2: How do our input representation as well as pre-training tasks contribute to FAIR's performance?}

For the input representation, we experiment with three variants of FAIR: (1) \textbf{FAIR w/o type}: remove the type information from all flows, i.e., only indicate whether the flow exists or not, (2) \textbf{FAIR w/o flow}: remove the bias in Equation~\ref{equation:attention_score} when calculating attention scores, and (3) \textbf{FAIR w/ CDFG}: replace the input with CDFG+call graph with the typed flow. With respect to pre-training tasks, we experiment with the following variants: (1) \textbf{FAIR w/o MLM}: remove the MLM pre-training task, (2) \textbf{FAIR w/o xFT}: remove the CFT/DFT pre-training tasks, (3) \textbf{FAIR w/o BVP}: remove the BVP pre-training task, (4) \textbf{FAIR w/o CL}: remove the contrastive learning pre-training task, and (5) \textbf{FAIR w/o all}: remove all pre-training tasks. The results are shown in Table~\ref{table:results_ablation}, where the worst results of each group are underlined.

Several observations deserve mention. First, each part of the input and each pre-training task can help FAIR to get better performance on downstream tasks. Second, for code-to-code retrieval and algorithm classification, changing the input to a CDFG has the greatest impact on performance, especially for GCJ. However, for heterogeneous device mapping, changing the input to a CDFG seems to have a smaller impact on performance. As for the contribution of the pre-training tasks, removing contrastive learning from the pre-training tasks has the biggest impact on the performance of the first two tasks. It is because contrastive learning enhances the model's capability to identify semantically similar and dissimilar IRs, which is what the model needs to perform well both downstream tasks. Finally, in most cases, for the last two tasks with very limited data, model performance does not show any significant change when we remove one of the pre-training tasks, but when we remove all of them, performance deteriorates.

\begin{table}[t]
    \centering
    \caption{Results on zero-shot code-to-code retrieval.}
    \label{table:results_rust}
    \resizebox{\linewidth}{!}{%
    \begin{tabular}{lrrrrrr}
        \toprule
            Models & CodeBERT & CodeT5 & UniXcoder & OSCAR & IRGen & FAIR  \\
        \midrule
            MAP@R  & 8.70    & 7.41  & 21.19     & 22.72 & 24.83 & 27.22 \\
        \bottomrule
    \end{tabular}
    }
\end{table} 

\subsection{Transferability}
\label{section:exp_results_rust}

\textbf{RQ3: How well can FAIR transfer to IR compiled from unseen programming languages in the zero-shot setting?}
\label{section:exp_eval_rq3}

We evaluate FAIR on the code-to-code retrieval tasks using a dataset of unseen programming languages. This experiment will also allow us to measure the ability of FAIR to represent the IR program of low-resource programming languages. Specifically, there are many niche or emerging languages that do not have the same active community and large-scale data as popular languages needed to effectively train a model. Although the source codes of programming languages share some lexical similarity, and existing work has demonstrated the ability of some source code-based models to transfer between programming languages, we believe that an IR-based approach is better suited to do this because IR can completely eliminate the differences between programming languages. 

We collect 10,751 Rust solutions to 59 online judge problems from the CodeNet Corpus~\cite{puri2021codenet}. The Rust program is compiled to LLVM IR by using Cargo 1.68.2\footnote{\url{https://doc.rust-lang.org/stable/cargo/}}. We only choose the pre-trained models in Section~\ref{section:exp_baselines} as baselines\footnote{Only pre-trained models can be evaluated in the zero-shot setting.}. Other settings are the same as those shown in Section~\ref{section:exp_downstream_cr}.  

Results are shown in Table~\ref{table:results_rust}. As can be seen, (1) FAIR achieves state-of-the-art performance, (2) the IR-based models (i.e., OSCAR, IRGen, and FAIR) are generally better than the source code-based models, and (3) the models with contrastive learning (i.e., UniXcoder, OSCAR, IRGen, and FAIR) have a significant advantage.

\subsection{Qualitative Error Analysis}

To understand the strengths and weaknesses of FAIR, we conduct a qualitative analysis of FAIR and two existing pre-trained models of IR (i.e., IRGen and OSCAR) and a method using CDFG (i.e., ProGraML). Specifically, we conduct an error analysis according to three groups of test examples taken from the POJ-104 dataset of the code-to-code retrieval task. The first group contains 50 examples randomly selected from all of the test examples that all of the four models handle correctly. The second group contains 50 examples randomly selected from all of the test examples for which FAIR is correct and the other three models are wrong. The third group contains 50 examples randomly selected from all the test examples for which none of the models handles correctly. We believe that this last group contains some of the most challenging examples.

By examining the examples in the first and second groups, we find that FAIR has strengths in handling IR programs with the following characteristics: 

(1) Longer IR programs: The average number of lines of IR programs in the first group is 243.26 (i.e., with 1083.34 tokens), while that in the second group is 256.84 (i.e., with 1336.02 tokens). This shows that FAIR performs better on longer IR programs, which can likely be attributed to the fact that we have scaled down the input size in FAIR. This also explains FAIR's bigger advantage on GCJ than on POJ-104 compared with the other models, and the significant performance degradation of GNNs-based GNN-CDFG and ProGraML on GCJ in Table ~\ref{table:results_cr_cls}. Recall that the IR of the code in GCJ is seven times longer than that of the code in POJ-104~\cite{li2022irgen}, but FAIR is able to scale down the size of the input IR program to be less affected by the increase in the size of the input.

(2) More functions: We find that in the first group, there is only one example with five or more functions, while the second group has eight. We speculate that our use of call graphs to connect the independent functions in the CFG and the DFG enables FAIR to get a better understanding of the relationships between functions. 

(3) More diverse opcodes: the average number of opcode types that a DFG has in each IR in the first group and second group are 12.68 and 16.32, respectively. This is because we explicitly assign the opcode information to the flow type, then use the self-attention bias and pre-training tasks to make the model learn this information. 

% A close examination of the third group of examples reveals the potential limitations of FAIR (and other models) in handling IR with more complex data types, especially data type-sensitive programs, such as those containing many data type conversions. 

A closer look at the third group of examples highlights FAIR's limitations in handling complex data types in IR, especially data type-sensitive programs with multiple conversions. This may be because we do not explicitly extract data types from the instruction nodes during DFG simplification, preventing the model from learning type-related information easily.

\section{Threats to Validity}
\label{section:threats}

\paragraph{Construct Validity} We do not check for duplicates in the pre-training data and the data on the downstream task, but we do not think this is a concern because the pre-training data contains neither algorithm-type data nor OpenCL programs, and therefore we think the impact of the data overlap on the downstream task is negligible. Prior studies follow the same setup~\cite{peng2021oscar,li2022irgen}.

\textit{External Validity.} We use the LLVM IR as the compiler intermediate representation since LLVM is one of the most popular compilers and supports lots of programming languages. We are not sure if our model has the same performance on other IRs such as the GCC IR. Previous work has chosen to use LLVM IR as well~\cite{ben2018ncc,venkatakeerthy2020ir2vec,cummins2021programl,li2022irgen}.

Besides, we evaluate the validity of FAIR on four tasks, including retrieval and classification tasks, with datasets containing IR compiled from C/C++ and OpenCL using Clang. We are not sure if FAIR will have different performance on other tasks, or on IR generated with other compiler front-ends. We used another programming language with a front-end in Section~\ref{section:exp_results_rust} to verify to some extent the external validity of FAIR at this point. Moreover, we make more choices of downstream tasks and datasets than in previous work, for example, Li et al.~\cite{li2022irgen} consider code-to-code retrieval on POJ-104 and GCJ, which is our first downstream task.

\section{Conclusion and Future Work}
\label{section:conclusion}

We proposed FAIR, a flow type-aware IR-based pre-trained model, which (1) reduces the input size and adds more flow type information by splitting the CDFG into a CFG and a DFG, simplifying the DFG, adding flow type information and call graph to the two graphs, and connecting a CFG and a DFG by adding flows to them; (2) uses the Transformer Encoder and Word Embedding to embed the nodes of a CFG and a DFG respectively and learn the flow information in the graph; and (3) employs five pre-training tasks to pre-train FAIR so that it can learn text semantics, flow information, and the overall representation of an IR program. By fine-tuning FAIR on four downstream tasks, we show that FAIR achieved state-of-the-art performance on all tasks. Our ablation study and zero-shot investigation experiment also demonstrated the advantages of the different components of FAIR and its representation capability.

In future work, we expect to use IR for representation learning at the project level since the compilation process can give more cross-file information and project-level information in the IR.

% In this paper, we only experiment on the IR on the program level, which is the IR file compiled from a single source file. However, we believe that the IR representation at the project level will also be advantageous because the compilation process can give more cross-file information and project-level information in the IR and graph information. We believe that IR-based and source-code-based models have different advantages and disadvantages for different tasks, and neither can completely replace the other. Therefore, it is also our future research direction to explore the advantages and disadvantages of the IR-based model and the source code-based model.

\begin{acks}
% This work is supported by the Cooperation Fund of Huawei-NJU Creative Laboratory for the Next Programming, CCF-Huawei Populus Grove Fund, NSF award 2034508, . We also thank the reviewers for their helpful comments. Chuanyi Li is the corresponding author.

This research / project is supported by the Cooperation Fund of Huawei-NJU Creative Laboratory for the Next Programming, CCF-Huawei Populus Grove Fund, NSF award 2034508, and the National Research Foundation, under its Investigatorship Grant (NRF-NRFI08-2022-0002). Any opinions, findings and conclusions or recommendations expressed in this material are those of the author(s) and do not reflect the views of National Research Foundation, Singapore. We also thank the reviewers for their helpful comments. Chuanyi Li is the corresponding author.
\end{acks}

\balance
\bibliographystyle{ACM-Reference-Format}
\bibliography{refs}

% \appendix

\end{document}